\documentclass[aps,twocolumn,showpacs,amsmath,amssymb,pre,superscriptaddress, floatfix,nofootinbib]{revtex4-1}

\usepackage{graphicx}
\usepackage{dcolumn}
\usepackage{bm}
\usepackage{amssymb}
\usepackage{multirow}

\newcommand{\1}{\begin{equation}}
\newcommand{\2}{\end{equation}}
\newcommand{\ea}{\begin{eqnarray}} 
\newcommand{\ee}{\end{eqnarray}}
\newcommand{\4}[2]{{\frac{#1}{#2}}}
 
\newcommand{\Sum}[2]{{\sum\limits_{#1}^{#2}}}

\begin{document}

\title{Patterned Deposition of Particles in Spatio-temporally Driven Lattices}

\date{\today}

\pacs{05.45.Gg,05.45.Ac}

\author{Benno Liebchen}
\email[]{Benno.Liebchen@physnet.uni-hamburg.de}
\affiliation{Zentrum f\"ur Optische Quantentechnologien, Universit\"at Hamburg, Luruper Chaussee 149, 22761 Hamburg, Germany}%
\author{Christoph Petri}
\affiliation{Zentrum f\"ur Optische Quantentechnologien, Universit\"at Hamburg, Luruper Chaussee 149, 22761 Hamburg, Germany}%
\author{Florian Lenz}
\affiliation{Zentrum f\"ur Optische Quantentechnologien, Universit\"at Hamburg, Luruper Chaussee 149, 22761 Hamburg, Germany}%
\author{Peter Schmelcher}
\email[]{Peter.Schmelcher@physnet.uni-hamburg.de}
\affiliation{Zentrum f\"ur Optische Quantentechnologien, Universit\"at Hamburg, Luruper Chaussee 149, 22761 Hamburg, Germany}%
\begin{abstract}
We present and analyze mechanisms for the patterned deposition of particles in a spatio-temporally driven lattice. 
The working principle is based on the breaking of the spatio-temporal translation symmetry, which is responsible 
for the equivalence of all lattice sites, by applying modulated phase shifts to the lattice sites. 
The patterned trapping of the particles occurs in confined chaotic seas, created via the ramping of the 
height of the lattice potential. Complex density profiles on the length scale of the complete lattice can be 
obtained by a quasi-continuous, spatial deformation of the chaotic sea in a frequency modulated lattice. 
\end{abstract}
\maketitle
\emph{Introduction}
The desire of understanding the fascinating mechanism that allows living organisms to extract mechanical energy from thermal fluctuations
evoked elaborate theoretical and experimental investigations \cite{hanggi09}.
A directed flow of particles can in principle arise from thermal fluctuations,
if an appropriate symmetry-breaking and suitable time-correlations are present \cite{magnasco93}.
The spatio-temporal symmetries to be broken lead to the existence of a counter-propagating partner to each trajectory
\cite{flach00}. Ratchet transport has consequently been shown to occur in the nonequilibrium dynamics of driven Hamiltonian systems 
with mixed phase space \cite{schanz01,denisov01,denisov07}.
A first experimental realization of an electron ratchet in a semiconductor nanostructure \cite{linke99} 
showed the invertibility of the direction of the current via the temperature.  
The full range of controllability of the ratchet-transport became clear in experiments in Josephson junction-arrays \cite{sterck02}
and especially in the remarkable realizations of ratchets with atoms in time-dependent optical lattices \cite{schiavoni03,gommers05,salger09}. 
On the theoretical side, most work has focused on the dynamics of a single barrier potential \cite{buttiker82} and on setups where the 
spatial and temporal dependencies of the underlying forces decouple \cite{flach00,denisov01,denisov02,schanz05}. 
It is only recently that a driven lattice with a spatially varying driving law has been theoretically explored \cite{petri10}
and the feasibility of a local engineering of phase space has been demonstrated.
Motivated by this enhanced control achieved in phase modulated lattices \cite{petri10}, we address here the question:
Is it possible to exploit spatio-temporally varying driving laws in order to realize a mechanism for selective trapping of particles in lattices thereby
leading to certain patterns of their spatial deposition?
\\In the present work, we demonstrate and analyze a mechanism that provides us with specific pattern-like distributions of particles in a lattice of periodically driven potential barriers, 
starting from a homogeneous initial distribution.
Considering a lattice with a certain broken spatio-temporal symmetry and ramping (i.e. instantaneously increasing) the lattice potential after a given time, we observe
that particles are trapped on specific sites of the unit cell, while the other sites are diffusively emptied.
Here, the periodicity of the resulting particle distribution follows the length scale of the unit cell. We also present a 
second mechanism for the creation of sophisticated particle distributions following certain globally varying spatial patterns.
\begin{figure}
\centering
\includegraphics[width=0.44\textwidth]{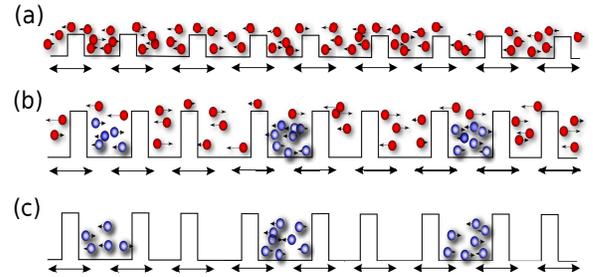} 
\caption{\small (Color online) Schematic illustration of the driven lattice and the trapping mechanism:
Before ramping the potential (a), after ramping (b) and at the end of the simulation (c).
Blue particles are trapped, red ones are diffusive.}
\label{schem}
\end{figure}

\emph{Setup}
We consider a system of noninteracting classical particles in a finite one-dimensional lattice of laterally oscillating potentials with equal height $V$,
described by the
following single particle Hamiltonian:
\begin{equation} \label{system} H = \4{p^2}{2m}+V \Sum{i=-N/2}{N/2} \theta \left( \4{l}{2}-\left| x-i L-a f(t) \right| \right); \end{equation}
Therein, $m$, $x$ and $p$ denote the mass, position and momentum of the particle and
$a$, $\omega$, $l$, $f(t)=\cos(\omega t +\phi_i)$ represent the amplitude, frequency, width and driving law of the oscillating barriers. $L$ is the equilibrium distance between two barriers and 
$\phi_i$ is the phase of the $i$-th barrier. A sketch of the lattice is shown in Fig. 1a. 
The space between the $i$-th and the $(i+1)$-st barrier will be referred to as the $i$-th lattice site. As a particle reaches the $\pm (N/2+1)$st barrier it is considered to be lost.
Obviously the system is unbiased, i.e. the time averaged net force vanishes.
With the scaling transformations
$t \mapsto t':=\omega t;\;\; x \mapsto x':= x/a$ and additionally
$\tilde H'(x',t'):=\4{m H}{\omega^2a^2}$ the number of parameters can be reduced by three.
Therefore, we may choose $m=l=1$ and $L=2+2a+l$ without loss of generality.

\emph{Spatio-temporal symmetry breaking}
What kind of spatio-temporal driving is necessary in order to achieve selective trapping into certain sites of a lattice?
Obviously, our aim cannot be achieved with a `global' (i.e. site independent) driving law, since then trajectories would obey the same dynamics on every lattice site.
In Ref. \cite{petri10} local changes in phase space were obtained in a lattice by employing linear phase shifts $\phi_i = i\cdot\phi$.
They are however not sufficient to lead to selective trapping of particles, since the following symmetry still holds: 
\1 \label{sym} H(x+L,t-\phi/\omega)=H(x,t) \2 
In lattices satisfying the symmetry (\ref{sym}) a trajectory which is at time $t$ at a position $x$
has the same future as a trajectory which is at $t-\phi/\omega$ at the position $x+L$.
That is, if particles are trapped on site $i$, those on site $i+1$ are also trapped.
For this reason, breaking the symmetry (\ref{sym}) of the Hamiltonian is a necessary requirement for selective trapping.
\\Following the above arguments we choose a system with a unit cell consisting of three barriers, i.e. $\phi_{i+3}=\phi_i$ and
nonlinear phase shifts $(\phi_0,\phi_1,\phi_2)=(0,\pi/10,3\pi/10)$.
Initially, our lattice consisting of $N=4000$ barriers, possesses a low potential height $V=1$ and we choose a uniformly distributed ensemble of $N_p=10^7$ identical, non-interacting classical particles
with randomly distributed, small velocities (corresponding to initial conditions in the chaotic sea of the underlying phase space).
The particle dynamics is followed until a system-specific long-time transient
and a quasi-stationary distribution in the lattice are reached.
Then at $t_{\rm r}:=2\times 10^4$ (which corresponds for each particle approximately to the same number of collisions with a barrier), the potential is ramped (instantaneously increased) to the larger value $V=9$ and the dynamics
is followed until $t_{\rm f}=10^6$.

\begin{figure}[htb]
\centering 
\includegraphics[width=0.44\textwidth]{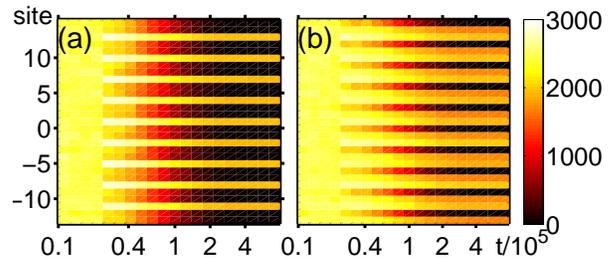}
\caption{\small (Color online) Representative extract of the time-evolution of the density of particles: The particle number (color) per site (density) is plotted versus
time (logarithmic scale) and site number. The potential is ramped at $t_{\rm r}=2\times 10^4$ from $V=1$ to $V=9$ (a) and to $V=15$ (b). ($a=\omega=1$).}
\label{colorpl}
\end{figure}

\begin{figure}[htb]
\centering
\includegraphics[width=0.44\textwidth]{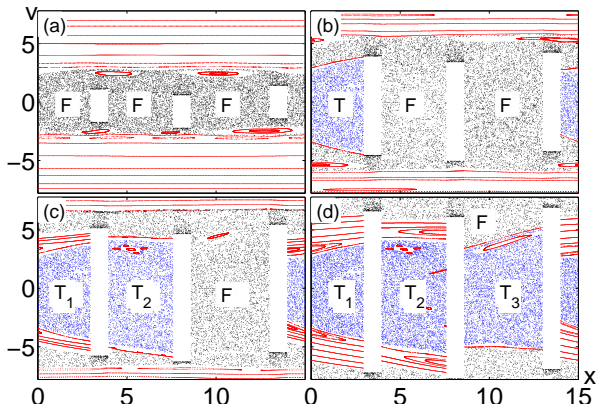}
\caption{\small (Color online) Stroboscopic Poincar\'{e} surfaces of section (PSOS) for the unit cell at times $t\;{\rm mod}\;2\pi=0$ for different values of the barrier potential 
$V=1,9,15,24$ for (a) to (d), respectively. Components corresponding to a localized chaotic sea (blue) are denoted by "T", those which are extended over the whole lattice by "F" (black).
Red curves show regular orbits. ($a=\omega=1$).
}
\label{poincsec}
\end{figure}

\begin{figure}[htb]
\centering 
\includegraphics[width=0.44\textwidth]{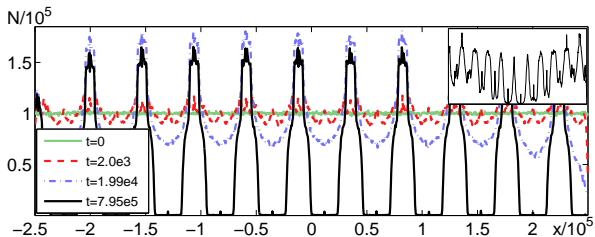}
\label{fig:bif0-7}
\caption{\small (Color online) Snapshots of the time-evolution of the particle-density in the lattice
with a spatially varying frequency $\omega_i=1+\sin(i/1500)$ (Inset: $\omega_i=1.2+0.25\sin(i/1500)-1.6 \cdot |i|/N$) ($a=1$).}
\label{globpat}
\end{figure}

\emph{Results}
Fig.~\ref{colorpl}a shows an extract of the time-evolution of the particle density in the lattice. 
Initially, the particle distribution is uniform.
After ramping the potential, we observe that the particle number remains approximately constant 
on every third lattice site, while the other sites are emptied (see Fig.~\ref{schem}b). Asymptotically, the distribution becomes stationary 
(see Fig.~\ref{schem}c) and only every third lattice site is occupied by particles whereas the in-between sites are empty. 
Analogously, Fig.~\ref{colorpl}b reveals, that ramping the potential from $V=1$ to $V=15$ results in trapping on two of three sites of the unit cell. 
\\We now turn to a quantitative discussion of the underlying process and address the questions of efficiency and predictability of the selective trapping mechanism.
Tab.~\ref{tabu} shows the final particle distribution at $t=t_{\rm f} \gg t_{\rm r}$ and
the first three columns list the fraction of particles (in percent), 
which are trapped on the first/second/third site of the unit cell,
taken with respect to the total number of particles which are in the lattice at $t=t_{\rm r}$.
We note, that trapping on the first site of the unit cell corresponds of course to trapping on the sites $i={..,-2,1,4,7,..}$ in the extended lattice.
Without ramping the potential, we observe from the $V=1$-row, that at $t=t_{\rm f}$ the lattice is completely emptied.
Ramping from $V=1$ to $V=9$ (to $V=15$/$V=24$), we see that the
fraction of particles trapped on the first site (the first two sites/all three sites) of the unit cell exceeds 
the fraction of particles trapped on the other (undesired) sites by about a factor 100. 
The small numbers of particles trapped on the undesired sites are a consequence of the impossibility
to distinguish long transient trapping from permanent trapping.
The second last column of Tab.~\ref{tabu} displays the overall fraction of particles which is trapped at $t_{\rm f}$. 
This number converges for high barrier potentials to the system specific limit of about $82\%$ of all available particles, shown in the case $V=100$.
Finally Tab.~\ref{tabu} shows the number of particles, which are neither trapped nor have left the lattice at the end of the simulation. 
Since this fraction of particles which are asymptotically diffusive is small, it is confirmed that the obtained pattern-like deposition becomes indeed stationary.
Note that all the relative numbers provided in Tab.~\ref{tabu} are independent of the specific initial particle distribution.
The ratio of the fluctuations (standard deviation) and the average number of particles per site is expected to
decrease asymptotically for large particle numbers, according to $1/\sqrt{N_p}$, with a system specific proportionality constant.
Our simulation of $10^7$ particles yields fluctuations of the particle number per site of $2.4\%$ for $V=9$.
\\In the following, we show that the mechanism of selective trapping can be traced back
to the Poincar\'{e} surface of section of the system. 
We firstly discuss how the breaking of the symmetry (\ref{sym}) leaves its hallmarks in the underlying phase space.
In Fig.~\ref{poincsec}a we observe that for the low barrier potential $V=1$, the chaotic sea
represents a connected area. Therefore, diffusion occurs all-over the lattice. 
For a higher barrier potential $V=9$, according to Fig.~\ref{poincsec}b a part of the chaotic sea is separated (region T) by regular spanning
curves from the main chaotic sea which corresponds to diffusion over the whole lattice (region F).
The particles in region T never leave the corresponding lattice site, i.e. they are trapped.
Simultaneously, particles in region F follow a diffusive motion until they reach the border of the lattice and leave it.
For a potential height $V=15$, Fig.~\ref{poincsec}c shows two isolated chaotic seas (regions T${_1}$, T${}_2$). Starting again with $V=1$ and ramping to $V=15$, trapping thus occurs on two sites
of each unit cell, as we have demonstrated in Fig.~\ref{colorpl}b. Finally, for $V=24$, the chaotic sea, (Fig.~\ref{poincsec}d) suggests trapping on all lattice-sites.
Every lattice site can be regarded as a Fermi-Ulam-model (FUM) \cite{fermi49,lichtenberg92} with finite wall potential. 
Accordingly, the trapping mechanism can be understood on a physically intuitive level:
In the FUM, due to the formation of correlations, there are regular spanning curves, which prohibit unlimited energy growth.
If the kinetic energy corresponding to the first invariant spanning curve (FISC) lies completely (i.e. for all phases of the oscillation and all coordinates) below the barrier potential height, particles below the barrier will not be
able to overcome it. 
In contrast to this, particles in the region F will escape from the lattice after some time.
In the investigated symmetry-broken lattice, the unit cell can be considered to consist of three different finite-wall-FUMs.
Thus, particles will be trapped on those sites of the unit cell which have FISCs which are completely below $V$.
Ramping the potential to $V=9$ ($V=15$), we encounter trapping in the unit cell on the site with the lowest (two lowest) FISC(s), since the FISC leads to a confined chaotic sea for the corresponding site(s).
\\According to Tab. \ref{tabu} even for a very large potential $V$ the number of trapped particles cannot exceed $82\%$.
This is due to the fact that particles which are at the location of a barrier at the moment when the potential is
increased, gain the potential energy $V(t>t_{\rm r})-V(t<t_{\rm r})$ and leave the lattice either on regular orbits or via diffusion above the FISC. 
Since the particle distribution for $t<t_{\rm r}$ is close to uniform,
the fraction of maximally trappable particles can be estimated as $L-l/L$.
For very small barriers $l \rightarrow 0$, the overall fraction of trappable particles is expected to approach $100\%$ which was confirmed by our numerical simulations.
The typical velocities of the FISC on a specific site strongly increases with the phase difference of the adjacent barriers for not too large values of this difference.
We can thus develop a procedure that yields a lattice which contains particles arranged according to a desired pattern.
We might for example wish as a final distribution to have particles only on every second lattice site.
To this end, a period-4-lattice with a unit cell consisting of two barriers with small and two barriers with large phase shifts serves the purpose.
For example, by choosing $(\phi_0,\phi_1,\phi_2,\phi_3)$ $=$ $(0,\4{3\pi}{4},\4{3\pi}{4},0)$ and ramping from $V=1$ to $V=100$, we obtain a final distribution of
(0.055/20.90/0.003/20.82; 41.77; 3.98) following the notation of Tab.~\ref{tabu}.
Also the relative occupation of the occupied sites can be changed by an additional modulation of the barrier amplitudes $a\rightarrow a_i;\quad a_i=a_{i+3};\quad i=0,1,2;$.
Choosing $a_1=2.0$; $a_2=0.5$; $a_3=1$ and ramping from $V=1$ to $V=100$ in the 
period-3-case, we obtain (1st/2nd/3rd; t.f.; d.f.)$=$(36.30/18.20/27.96; 82.44; 0.20) and for $a_1=0.5$; $a_2=2.0$; $a_3=1$: (23.83/37.98/20.69; 82.47; 0.20).
This demonstrates the tunability of the relative occupation probability of the lattice sites.
\\The mechanism of patterned deposition is expected to be stable with respect to small thermal fluctuations of the ensemble of particles, because the breaking of the symmetry (\ref{sym}) is not affected by the latter. 
However they will cause a certain transmissibility of the spanning curves, because the energy change provided by the thermal fluctuations enables particles to overcome the spanning curve.
As more and more spanning curves occur with increasing $V$, this leaking rate can be tuned to an arbitrary small value.
\\Since the ratchet effect can occur for both, particles in spatio-temporally-driven potentials, and thermally driven particles in a static potential, it is interesting to ask whether the
mechanism of patterned deposition can also be obtained in thermally driven setups.
For a system, consisting of two species, an enormous flexibility, concerning direction and magnitude of the ratchet current was demonstrated \cite{savelev03, savelev04, potosky10}, thus suggesting that periodic clustering can also occur 
in a thermally driven setup.
\begin{center}
\begin{table}
\caption{\label{tabu} 
Propagation of $10^7$ particles, for $t=10^6$ time steps in a lattice consisting of 4000 individual barriers (parameters: Fig.~\ref{colorpl}).
At the ramping time $t_{\rm r}=2\times 10^4$, $19.64\%$ of all $10^7$ particles have left the lattice.
First column: Barrier-potential $V(t>t_{\rm r})$. 
"site 1(2,3)": Conditional probability
for a particle which was still in the lattice at $t_{\rm r}$, to be trapped
on the first (second, third) site of the unit cell at $t=t_{\rm f}$.
"total fraction": Fraction of trapped particles, related to the total number of particles which are in the lattice
at $t_{\rm r}$. The total fraction of diffusive particles in per cent at $t=10^6$ is also shown.}
\begin{ruledtabular} 
\begin{tabular}{|l|l|l|l|l|l|}\hline
$V$ & site 1 & site 2 & site 3 & total fraction & diffusive fraction \\\hline
1 & 0.00 & 0.00 & 0.00 &  0.00 & 0.36 \\\hline
9 & 26.54 & 0.15 & 0.26 & 26.95 & 0.38 \\\hline
15 & 26.66 & 23.38 & 0.27 & 50.32 & 0.37 \\\hline
24 & 26.66 & 23.38 & 27.67  & 77.71 & 0.77 \\\hline
100 & 26.66 & 23.38 & 31.93 & 81.97 & 0.37 \\\hline
\end{tabular}
\end{ruledtabular}
\end{table}
\end{center}
\emph{Global patterns of particle deposition}
We explore now the question, whether a selective trapping of particles is possible,
on the length scale of the complete lattice, i.e. following patterns beyond the size of a few-site unit-cell.
It turns out, that a frequency-modulated lattice is the key to achieve global modulations of the density. 
Let us consider a uniformly distributed ensemble of $10^7$ particles with initially low, randomly distributed velocities in a lattice consisting of $N=10^5$ barriers possessing a site dependent frequency
$\omega \rightarrow \omega_i := 1+\sin(i/1500)$. We integrate 
the dynamics for $V=1$ up to $t_{\rm r}=2\times 10^4$ and subsequently ramp the barriers to $V=120$, followed by a further propagation to $t_{\rm f}=10^6$.
\\Fig.~\ref{globpat} shows the corresponding evolution of the particle density at four different times.
Initially, the ensemble is uniformly distributed. For $t \ll t_{\rm r}$ 
the emergence of a (quasi-) periodic global density modulation on large length scales is conspicuous. The peaks develop in those regions of the lattice where the 
barrier frequency is lower. For $t\lesssim t_{\rm r}$ the observed modulation is more pronounced. 
Finally, for $t \gg t_{\rm r}$, the high-frequency regions of the lattice are completely emptied.
The resulting density distribution has become stationary.
\\Let us investigate the mechanism for the time-evolution of the particle density
leading to a localization of the
 particle distribution in the regions of the lattice with slowly oscillating barriers.
Here, we consider the behavior of the FISC on a specific lattice site.
In the high frequency-regions of the lattice, the FISC occurs at higher velocities than in the low-frequency regions. 
This leads to two important effects:
Firstly, we have a relatively high average particle velocity in the high frequency-regions, corresponding
to a short average dwell time and resulting in a low particle density when compared to regions with slowly oscillating barriers. 
Since the barrier frequency varies quasi-periodically over the lattice, the resulting particle density varies likewise with the same periodicity.
Secondly, the maximal kinetic energy a particle can achieve on a lattice site increases with the typical velocity of the FISC. 
Thus, after ramping the potential to an appropriate value, following the above discussion of the selective localization in a few-site unit cell, 
particles will be trapped on those sites, where the FISC is located completely below the potential height. 
The latter occurs in the low-frequency regions of the driven lattice.
\\On basis of the above-developed understanding, we can now further design particle distributions by appropriately choosing the barrier frequencies:
For instance, the final particle distribution for the superposition of a periodic frequency-modulation and a piecewise constant gradient, declining from the center of the lattice to both edges $\omega_i=1.2+0.25\sin(i/1500)-1.6 \cdot |i|/N$,
is shown in the inset of Fig.~\ref{globpat}. 
The density reflects the expectation of a particle accumulation in the low-frequency regions.
Note that for the continuous manipulation of the FISC it is necessary to have a small frequency change from barrier to barrier. 
Otherwise, particles in the fast regions could move on regular orbits traversing the low frequency regions, instead of being captured in the chaotic sea in the course of the ramp.

\emph{Conclusion}
We have demonstrated and analyzed mechanisms for the patterned deposition of particles in a lattice by applying spatially varying driving laws and a corresponding ramp of the potential height. 
The symmetry which leads to the long-term equivalence of all lattice sites has been broken by appropriately chosen phase shifts of the barriers. This results in
trapping in a patterned series of confined chaotic seas. 
Employing a frequency modulated lattice we demonstrated that even patterned density profiles on the length scale of the complete lattice can be achieved.
A possible experimental setup to demonstrate the mechanisms of selective trapping is provided by cold atoms in optical lattices.
Oscillating mirrors and acousto-optic modulators can be used to obtain a laterally oscillating potential landscape. Superimposing laser beams of 
different frequency (amplitude) and adjusting the relative phases by controlling the positions of the light sources on a nm-scale we expect to arrive at setups which show the demonstrated
mechanism of patterned deposition. The ramp leading to the final confinement in the individual sites can be implemented by an instantaneous switch of the amplitude of the laser(s).
An alternative physical system to realize our model is provided by multi-layered semiconductor heterostructures, where applied AC voltages 
can be used to drive each layer individually.
Potential applications include selective loading of a lattice with cold atoms, thus permitting local addressing for quantum information processing. 
It might be also of relevance to atom lithography.

\paragraph*{Acknowledgments}
We thank F.K.~Diakonos for useful discussions.

\end{document}